%% file: zendk.tex
\documentclass[submission,copyright,creativecommons]{eptcs}
\providecommand{\event}{PxTP~2015} 
\usepackage{breakurl}             

\usepackage[T1]{fontenc}
\usepackage{bussproofs}
\usepackage{url}
\usepackage{amsmath,amssymb}
\usepackage{graphicx}
\usepackage{tabularx}
\usepackage{multirow}
\input{macros}
\DeclareSymbolFont{letters}{OML}{txmi}{m}{it}

\title{Checking \zenm{} Proofs in \dedukti{}
  \thanks{This work has received funding from the \bware{} project
    (ANR-12-INSE-0010) funded by the INS programme of
    the French National Research Agency (ANR).}
}
\author{Rapha\"{e}l Cauderlier\quad{}\quad{}Pierre Halmagrand
  \institute{\cnam{} - \inria{}\\Paris, France}
  \email{raphael.cauderlier@inria.fr\quad{}pierre.halmagrand@inria.fr}}

\def\titlerunning{Checking \zenm{} Proofs in \dedukti{}}
\def\authorrunning{R. Cauderlier \& P. Halmagrand}
\begin{document}
\maketitle

\begin{abstract}
  \dedukti{} has been proposed as a universal proof checker. It is a
  logical framework based on the $\lambda\Pi$-calculus modulo that is used
  as a backend to verify proofs coming from theorem provers,
  especially those implementing some form of
  rewriting. We present a shallow embedding into \dedukti{} of proofs
  produced by \zenm{}, an extension of the tableau-based first-order
  theorem prover \zenon{} to deduction modulo and typing. \zenm{}
  is applied to the verification of programs in both academic and
  industrial projects. The purpose of our embedding is to increase
  the confidence in automatically generated proofs by separating untrusted
  proof search from trusted proof verification.
\end{abstract}

\input{intro}
\input{modul}
\input{deduk}

\input{zenon}
\input{bench}
\input{concl}

\bibliographystyle{eptcs}
\bibliography{biblio}

\clearpage
\appendix
\input{appnd}

\end{document}

%% file: macros.tex
\def\altergo{\textsf{Alt-Ergo}}

\def\bmth{\textsf{B}}
\def\bware{\textsf{BWare}}
\def\focalize{\textsf{FoCaLiZe}}
\def\focalide{\textsf{Focalide}}

\def\cnam{\textsf{Cnam}}

\def\dedukti{\textsf{Dedukti}}

\def\inria{\textsf{Inria}}

\def\iproverm{\textsf{iProver~Modulo}}
\def\hol{\textsf{HOL}}

\def\why{\textsf{Why3}}
\def\whyml{\textsf{WhyML}}
\def\zenm{\textsf{Zenon~Modulo}}
\def\zenon{\textsf{Zenon}}

\def\z3{\textsf{Z3}}
\def\boogie{\textsf{Boogie}}
\def\llproof{\textsf{LLproof}}
\def\coq{\textsf{Coq}}
\def\ocaml{\textsf{OCaml}}
\def\tff{\textsf{TFF1}}

\EnableBpAbbreviations

\def\fCenter{\mbox{$\ \vdash\ $}}

\def\Type{\textsf{Type}}
\def\Kind{\textsf{Kind}}
\def\Prop{\textsf{Prop}}
\def\proof{\textsf{prf}}
\def\type{\textsf{type}}
\def\term{\textsf{term}}

\def\arr{\rightarrow}
\def\arg{\mbox{-}}
\def\dkrew{\hookrightarrow}
\newcommand{\mrule}{\textsf{R}}
\newcommand{\seq}{\ensuremath{\proof~\bot}}

\def\imp{\Rightarrow}
\def\eqv{\Leftrightarrow}

\def\set{\mathsf{set}}

\def\omicron{o}
\newcommand{\rew}{\ensuremath{\mathbin{\longrightarrow}}}
\newcommand{\rewdelta}{\ensuremath{\mathbin{{\longrightarrow}_\Delta}}}

\def\fctpart{~{\kern+.2em\mapstochar{}\!\!\!\!\!\rightarrow}~}

\def\surpart{~{\kern+.2em\mapstochar{}\!\!\!\!\!\twoheadrightarrow}~}
\def\bijpart{~{\kern+.1em\mapstochar{}\!\!\!\!\!\rightarrowtail\kern+.03em
\!\!\!\!\!\!\twoheadrightarrow}~}
\def\bijtotal{~{\kern-.1em\rightarrowtail\kern+.03em\!\!\!\!\!\!
\twoheadrightarrow}~}

\def\dashedrule#1#2#3{{%
\dimen1=#2 \divide\dimen1 by 2
\def\@ruledash{%
\rule{\dimen1}{0pt}%
\rule[0.5ex]{#1}{0.4pt}%
\rule{\dimen1}{0pt}}%
\count1=0
\loop%
\ifnum\count1<#3%
\advance\count1 by 1%
\@ruledash%
\repeat}}

\newlength\memparindent
\newcommand{\fst}[1]{\mathsf{fst}(#1)}
\newcommand{\snd}[1]{\mathsf{snd}(#1)}
\newcommand{\pair}[2]{#1, #2}

\newcommand{\tr}[1]{\ensuremath{\left| #1 \right|}}
\newcommand{\dtr}[1]{\ensuremath{\| #1 \|}}

\newcommand{\figitem}[1]{\hspace{0.2cm}#1}

\newcommand{\myfigure}[3]
{
\begin{figure}[ht!]
  \framebox[\textwidth][c]
    {\parbox{\textwidth}{\small
        #3
    }}
 \caption{#1}
 \label{#2}
\end{figure}}

\newcommand{\wellformed}{\vdash}
\newcommand{\emptycontext}{\emptyset}

\newcommand{\bool}{\mathsf{bool}}
\newcommand{\btrue}{\mathsf{true}}
\newcommand{\bfalse}{\mathsf{false}}
\newcommand{\band}{~\&\&~}
\newcommand{\bor}{~||~}
\newcommand{\bnot}{\sim}
\newcommand{\ifte}[4]{\mathsf{if}_{#1}~#2~\mathsf{then}~#3~\mathsf{else}~#4}

\newcommand{\T}{\ensuremath{\mathcal{T}}}
\newcommand{\name}{\ensuremath{\mathrm{name}}}
\newcommand{\LPm}{\ensuremath{\lambda\Pi^{\equiv}}}

\newcommand{\tffm}{\ensuremath{\mathsf{TFF1}^\equiv}}
\newcommand{\llproofm}{\llproof$^\equiv$}

%% file: intro.tex
\section{Introduction}

Program verification using deductive methods has become a valued
technique among formal methods, with practical applications
in industry. It guarantees a high level of confidence regarding the
correctness of the developed software with respect to its
specification. This certification process is generally based on the
verification of a set of proof obligations, generated by
deductive verification tools. Unfortunately, the number of proof
obligations generated may be very high. To address this issue,
deductive verification tools often rely on automated deduction tools
such as first-order Automated Theorem Provers (ATP) or Satisfiability Modulo
Theories solvers (SMT) to automatically discharge a large number
of those proof obligations. For
instance, \boogie{} is distributed with the SMT
\z3~\cite{barnett2006boogie} and the \why{} platform with
\altergo~\cite{FP13}. After decades of constant work, ATP and SMT
have reached a high level of efficiency and now discharge more proof obligations
than ever. At the end, many of these program verification
tools use their corresponding ATP or SMT as oracles. The main concern
here is the level of confidence users give to them. These programs are
generally large
software, consisting of dozens of thousands of lines of code, and using some
elaborate heuristics, with some ad hoc proof traces at best,
and with a simple ``yes or no'' binary answer at worst.

A solution, stated by Barendregt and Barendsen~\cite{BB02} and
pursued by Miller~\cite{miller2011proposal} among others, relies on
the concept of proof certificates. ATP and SMT should be seen as
proof-certificate generators. The final ``yes or no'' answer is therefore
left to an external proof checker. In addition, Barendregt and
Barendsen proposed that proof checkers should satisfy two principles
called the De Bruijn criterion and the Poincar\'{e} principle.
The former states that proof checkers have to be built
on a light and auditable kernel. The latter recommends
that they distinguish reasoning and computing and that it should not be
necessary to record pure computational steps.

Relying on an external proof checker to verify proofs strongly
increases the trust we give them, but it also provides a common
framework to express proofs. A profit made by using
this common framework is the possibility to share proofs coming
from different theorem provers, relying on different proof systems. But nothing
comes for free, and using the same proof checker does not guarantee
in general that we can share proofs because formul\ae{} and proofs
can be translated in incompatible ways.
Translation of proofs must rely on a
shallow embedding in the sense proposed by Burel~\cite{burel2013shallow}:
it reuses the features of the target language. It does not introduce
new axioms and constants for logical symbols and inference rules.
Connectives and binders of the underlying logic of ATP
are translated to their corresponding connectives and binders
in the target language. In addition, a shallow embedding preserves the
computational behavior of the original ATP and the underlying
type system of the logic.

In this paper, we present a shallow embedding of \zenm{} proofs into
the proof checker \dedukti{}, consisting of an encoding of a typed
classical sequent calculus modulo into the $\lambda\Pi$-calculus modulo
(\LPm{} for short). \zenm~\cite{Zenon-Modulo} is an extension to
deduction modulo~\cite{DA03} of the first-order tableau-based ATP
\zenon~\cite{Zenon}. It has also been extended to support ML
polymorphism by implementing the \tff{} format~\cite{TFF1}.
\dedukti~\cite{BA12} is a proof checker that implements \LPm{}, a
proof language that has been proposed as a proof standard for proof
checking and interoperability.  This embedding is used to certify
proofs in two different projects: \focalize~\cite{focalize}, a
programming environment to develop certified programs and based on a
functional programming language with object-oriented features, and
\bware~\cite{BWare}, an industrial research project that aims at
providing a framework for the automated verification of proof
obligations coming from the \bmth{} method~\cite{B-Book}.  The main
benefit of \zenm{} and \dedukti{} relies on deduction
modulo. Deduction modulo is an extension of first-order logic that
allows reasoning modulo a congruence relation over propositions. It is
well suited for automated theorem proving when dealing with theories
since it turns axioms into rewrite rules.  Using rewrite rules during
proof search instead of reasoning on axioms lets provers focus on the
challenging part of proofs, speeds up the tool and reduces the size of
final proof trees~\cite{Zenon-Modulo2}.

\zenon{} was designed to support \focalize{} as its dedicated deductive
tool and to generate proof certificates for \coq{}. Extension to deduction
modulo constrains us to use a proof checker that can easily reason modulo
rewriting. \dedukti{} is a good candidate to meet this specification.
A previous embedding of \zenm{} proofs into \dedukti{}, based on a
$\neg\neg$ translation~\cite{Zenon-Modulo}, was implemented as a tool
to translate classical proofs into constructive ones. This tool has the
benefit to be shallower since it does not need add the excluded
middle as an axiom into the target logic defined in \dedukti{}, but
in return this transformation may be very time-consuming~\cite{Zenon-Modulo2}
and was not
scalable to large proofs like those produced in \bware{}.
The closest related work is the shallow embedding of resolution
and superposition proofs into \dedukti{} proposed by
Burel~\cite{burel2013shallow} and implemented in \iproverm~\cite{GB11b}.
Our embedding is close enough to easily share proofs of
\zenm{} and \iproverm{} in \dedukti{}, at least for the subset of
untyped formul\ae{}.

The first contribution presented in this paper consists in the
encoding into \LPm{} of typed deduction modulo and a set of translation
functions into \LPm{} of theories expressed in this logic.
Another contribution of this paper is the extension to deduction modulo
and types of the sequent-like proof system \llproof{} which is the output
format of \zenm{} proofs.
The latter contribution is the embedding of this proof system into \LPm{}
and the associated translation function for proofs coming from this
system.

This paper is organized as follows: in Sec.~\ref{sec:modulo}, we
introduce typed deduction modulo; in Sec.~\ref{sec:dedukti},
we present \LPm{}, its proof checker \dedukti{}, and a
canonical encoding of typed deduction modulo in
\LPm{}; Sec.~\ref{sec:zenonmodulo} introduces
the ATP \zenm{}, the proof system \llproof{} used by \zenm{} to output
proofs; and the translation scheme implemented as the new output of
\zenm{}; finally, in Sec.~\ref{sec:bench}, we present some examples
and results to assess our implementation.

%% file: modul.tex
\section{Typed Deduction Modulo}
\label{sec:modulo}

The Poincar\'{e} principle, as stated by Barendregt and
Barendsen~\cite{BB02}, makes a distinction between deduction and
computation. Deduction may be defined using a set of inference rules
and axioms, while computation consists mainly in
simplification and unfolding of definitions. When dealing with axiomatic
theories, keeping all axioms on the deduction side leads to inefficient
proof search since the proof-search space grows with the theory.
For instance, proving the following statement:
\begin{align*}
  \fst{\pair a a} &= \snd{\pair a a}
\end{align*}
where $a$ is a constant, and \textsf{fst} and \textsf{snd} are defined by:
\begin{align*}
  \forall{}x,y.\;\fst{\pair x y}=x \hspace{2.2cm}
  \forall{}x,y.\;\snd{\pair x y}=y
\end{align*}
and with the reflexivity axiom:
\begin{align*}
  \forall{}x.\;x&=x
\end{align*}
using a usual automated theorem proving method such as tableau, will
generate some useless boilerplate proof steps, whereas a simple
unfolding of definitions of \textsf{fst} and \textsf{snd} directly
leads to the formula $a=a$.

Deduction modulo was introduced by Dowek, Hardin and
Kirchner~\cite{DA03} as a logical formalism to deal with axiomatic
theories in automated theorem proving. The proposed solution is to remove
computational arguments from proofs by reasoning modulo a decidable
congruence relation $\equiv$ on propositions. Such a congruence may
be generated by a confluent and terminating system of rewrite rules
(sometimes extended by equational axioms).

In our example, the two definitions may be replaced by the rewrite
rules:
\begin{align*}
  \fst{\pair x y}\rew{}x \hspace{2.2cm} \snd{\pair x y}\rew{}y
\end{align*}
And we obtain the following equivalence between propositions:
\begin{align*}
\left(\fst{\pair a a}=\snd{\pair a a}\right) \; &\equiv{} \;
\left(a=a\right)
\end{align*}

Reasoning with several theories at the same time is often necessary
in practice. For instance, in the \bware{} project, almost all proof
obligations combine the theory of booleans, arithmetic and set theory.
In this case, we have to introduce an expressive enough type system to
ensure that an axiom about booleans, for instance
$\forall{}x.~x=\mathsf{true}\lor{}x=\mathsf{false}$, will not
be used with a term that has another type. An input format for ATP called
\tff{}~\cite{TFF1} has been proposed recently by Blanchette and Paskevich
to deal with first-order problems with polymorphic types. We propose to
extend this format to deduction modulo.

We now introduce the notion of typed rewrite system, extending notations
of Dowek et al.~\cite{DA03}. In the following, $\mathrm{FV}(t)$ stands for the set
of free variables of $t$ where $t$ is either a \tff{} term or a \tff{} formula. \\

\begin{figure}[ht!]
  \framebox[\textwidth][c] {\parbox{\textwidth} {\small
               \begin{center}
                 \begin{tabular}{lr@{\hspace{1pt}}c@{\hspace{1pt}}l
                   @{\hspace{1.0cm}}l}
                   Types &
                   $\tau$ & $~::=~$ & $\alpha{}$ & (type variable)\\
                   & & $|$ & $T(\tau_1,\ldots{},\tau_m)$ & (type constructor)\\
                   Terms &
                   $e$ & $~::=~$ & $x$ & (term variable)\\
                   & & $|$ & $f(\tau_1,\ldots{},\tau_m;e_1,\ldots{},e_n)$ &
                   (function)\\
                   Formul\ae{} &
                   $\varphi$ & $~::=~$ & $\top~|~\bot$ & (true, false)\\
                   & & $|$ & $\neg{}\varphi~|~\varphi_1\land{}\varphi_2~|~
                   \varphi_1\lor{}\varphi_2~|~\varphi_1\imp{}\varphi_2
                   ~|~\varphi_1\eqv{}\varphi_2$ & (logical connectors)\\
                   & & $|$ & $e_1=_{\tau}e_2$ & (term equality)\\
                   & & $|$ & $P(\tau_1,\ldots,\tau_m;e_1,\ldots,e_n)$ &
                   (predicate)\\
                   & & $|$ & $\forall{}x:\tau.~\varphi(x)~|~\exists{}x:\tau.~
                   \varphi(x)$ & (term quantifiers)\\
                   & & $|$ & $\forall_{\type}\alpha:\type.~\varphi(\alpha)~|~
                   \exists_{\type}\alpha:\type.~\varphi(\alpha)$ &
                   (type quantifiers)\\
                   Context &
                   $\Delta$ & $~::=$ & $\emptyset$ & (empty context) \\
                   & & $|$ & $\Delta,x:\tau$ & (declaration) \\
                   Theory &
                   $\T$ & $~::=~$ & $\emptyset$ & (empty theory) \\
                   & & $|$ & $\T,T/m$ & (m-ary type constructor declaration\\
                   & & $|$ & $\T,f:\Pi\vec{\alpha}.~
                   \vec{\tau} \arr \tau$ & (function declaration)\\
                   & & $|$ & $\T,P:\Pi\vec{\alpha}.~
                   \vec{\tau} \arr \omicron$ & (predicate declaration)\\
                   & & $|$ & $\T,\name:\varphi$ & (axiom) \\
                   & & $|$ & $\T,l\rewdelta{}r$ & (rewrite rule)
                 \end{tabular}
               \end{center}
             } }
           \caption{Syntax of \tffm{}}
           \label{fig:syntax}
\end{figure}

\noindent\textbf{Definition (Typed Rewrite System)}\\ A term rewrite rule is
a pair of \tff{} terms $l$ and $r$ together with a \tff{} typing context $\Delta$
denoted by $l\rewdelta{}r$,
where $\mathrm{FV}(r)\subseteq\mathrm{FV}(l)\subseteq\Delta$. It is well-typed
in a theory $\T$ if $l$ and $r$ can be given the same type $A$ in $\T$ using
$\Delta$ to type free variables.
A proposition rewrite rule is
a pair of \tff{} formul\ae{} $l$ and $r$ together with a typing context $\Delta$
denoted by $l\rewdelta{}r$, where $l$ is an atomic
formula and $r$ is an arbitrary formula, and where
$\mathrm{FV}(r)\subseteq\mathrm{FV}(l)\subseteq\Delta$. It is well-typed in a
theory $\T$ if both $l$ and $r$ are well-formed formul\ae{} in $\T$ using
$\Delta$ to type free variables.

A typed rewrite system is a set $\mathcal{R}$ of proposition rewrite
rules along with a set $\mathcal{E}$ of term rewrite rules.  Given a
rewrite system $\mathcal{RE}$, the relation $=_\mathcal{RE}$ denotes
the congruence generated by $\mathcal{RE}$.  It is
well-formed in a theory $\T$, if all its rewrite rules are well-typed
in $\T$.

The notion of \tff{} theory can be extended with rewrite rules;
we call the resulting logic \tffm{}.
Its syntax is given in Fig.~\ref{fig:syntax}.

%% file: deduk.tex
\section{\dedukti{}}
\label{sec:dedukti}

The $\lambda\Pi$-calculus~\cite{barendregt2013lambda} is the simplest Pure
Type System featuring dependent types. It is commonly used as a
logical framework for encoding
logics~\cite{Harper1993}. The $\lambda\Pi$-calculus modulo, presented in
Fig.~\ref{fig:lpm}, is an extension of the $\lambda\Pi$-calculus with
rewriting. The $\lambda\Pi$-calculus modulo (abbreviated as \LPm{}) has
successfully been used to encode many logical systems
(\coq{}~\cite{boespflug2012coqine}, \hol{},
\iproverm~\cite{burel2013shallow}, \focalize) using shallow
embeddings.

In \LPm, conversion goes beyond simple $\beta$-equivalence since it
is extended by a custom rewrite system. When this rewrite system
is both strongly normalizing and confluent, each term gets a unique
(up to $\alpha$-conversion) normal form and both conversion and
type-checking become decidable. \dedukti{} is an implementation of
this decision procedure.

\myfigure{The $\lambda\Pi$-calculus modulo}{fig:lpm}{
  \figitem {Syntax}

  \begin{center}
    \begin{tabular}{lll}
      $s$ & $::=$ & $\Type$ $|$ $\Kind$\\
      $t$ & $::=$ & $x$ $|$ $t~t$ $|$ $\lambda x : t. t$ $|$ $\Pi x : t. t$ $|$ $s$ \\
      $\Delta$ & $::=$ & $\emptycontext$ $|$ $\Delta, x : t$ \\
      $\Gamma$ & $::=$ & $\emptycontext$ $|$ $\Gamma, x : t$ $|$ $\Gamma, t \dkrew_\Delta t$
    \end{tabular}
  \end{center}

  \figitem {Well-formdness}

  \begin{center}
    \begin{tabular}{ccc}
      \AxiomC{}
      \RightLabel {(Empty)}
      \UIC {$\emptycontext \wellformed$}
      \DisplayProof
      &
        \AxiomC{$\Gamma \wellformed$}
        \AxiomC{$\Gamma \vdash A : s$}
        \AxiomC{$x \notin \Gamma$}
        \RightLabel {(Decl)}
        \TIC {$\Gamma, x : A \wellformed$}
        \DisplayProof
      &
        \AxiomC{
           \begin{tabular}{l}
             $\Gamma, \Delta \vdash l : A$\\
             $\Gamma, \Delta \vdash r : A$
           \end{tabular}
         }
        \AxiomC{
           \begin{tabular}{l}
             $\Gamma, \Delta \vdash A : \Type$\\
             $FV(r) \subseteq FV(l) \subseteq \Delta$
           \end{tabular}
         }
        \RightLabel {(Rew)}
        \BIC {$\Gamma, l \dkrew_\Delta r \wellformed$}
        \DisplayProof
    \end{tabular}
  \end{center}

  \figitem {Typing}

  \begin{center}
    \begin{tabular}{cc}
      \AxiomC{$\Gamma \wellformed$}
      \RightLabel {(Sort)}
      \UIC {$\Gamma \vdash \Type : \Kind$}
      \DisplayProof
      &
        \AxiomC{$\Gamma \wellformed$}
        \AxiomC{$x : A \in \Gamma$}
        \RightLabel {(Var)}
        \BIC {$\Gamma \vdash x : A$}
        \DisplayProof
      \\ & \\
      \AxiomC{$\Gamma \vdash t_1 : \Pi x : A. B(x)$}
      \AxiomC{$\Gamma \vdash t_2 : A$}
      \RightLabel {(App)}
      \BIC {$\Gamma \vdash t_1~t_2 : B(t_1)$}
      \DisplayProof
      &
        \AxiomC{$\Gamma \vdash A : \Type$}
        \AxiomC{$\Gamma, x : A \vdash t : B(x)$}
        \AxiomC{$\Gamma, x : A \vdash B(x) : s$}
        \RightLabel {(Abs)}
        \TIC {$\Gamma \vdash \lambda x : A. t(x) : \Pi x : A. B(x)$}
        \DisplayProof
      \\ & \\
      \AxiomC{$\Gamma \vdash A : \Type$}
      \AxiomC{$\Gamma, x : A \vdash B(x) : s$}
      \RightLabel {(Prod)}
      \BIC {$\Gamma \vdash \Pi x : A. B(x) : s$}
      \DisplayProof
      &
        \AxiomC{$\Gamma \vdash t : A$}
        \AxiomC{$\Gamma \vdash B : s$}
        \AxiomC{$A \equiv_{\beta\Gamma} B$}
        \RightLabel {(Conv)}
        \TIC {$\Gamma \vdash t : B$}
        \DisplayProof
    \end{tabular}
  \end{center}
}

Burel~\cite{burel2013shallow} defines two encodings of deduction modulo in
\dedukti{}: a deep encoding $\tr \varphi$ in which logical connectives
are simply declared as \dedukti{} constants and a shallow encoding
$\dtr \varphi := \proof~{\tr \varphi}$ using a decoding function
$\proof$ for translating connectives to their impredicative encodings.
In Sec.~\ref{subsec:deepembedding} and
Sec.~\ref{subsec:deep_to_shallow}, we extend these encodings to
\tffm{}.

\subsection{Deep Embedding of Typed Deduction Modulo in \dedukti{}}
\label{subsec:deepembedding}

In Fig.~\ref{fig:deeplogic1}, for each symbol of our first-order typed
logic, we declare its corresponding symbol into \LPm{}.  In \LPm{},
types cannot be passed as arguments (no polymorphism) so we have to
translate \tffm{} types as \dedukti{} terms. The \dedukti{} type of
translated \tffm{} types is $\type$ and we can see an inhabitant of
$\type$ as a \dedukti{} type thanks to the $\term$ function.

In Fig.~\ref{fig:trfun}, we define a direct translation of \tffm{} in
\dedukti{}. It is correct in the following sense:
\begin{itemize}
\item if the theory $\T$ is well-formed in \tffm{}, then $\tr \T \wellformed$.
\item if $\tau$ is a well-formed \tffm{} type in a theory \T, then $\tr \T \vdash \tr \tau : \type$.
\item if $t$ is a \tffm{} term of type $\tau$ in a theory \T, then $\tr \T \vdash \tr t : \term~\tr \tau$.
\item if $\varphi$ is a well-formed \tffm{} formula in a theory \T, then $\tr \T \vdash \tr \varphi : \Prop$.
\end{itemize}

\begin{figure}[ht!]
  \framebox[\textwidth][c]
           {\parbox{\textwidth}
             {\small
               \hspace{0.2cm}Primitive Types
               \begin{center}
                 \begin{tabular}{r@{}l@{\hspace{1cm}}r@{}l
                     @{\hspace{1cm}}r@{}l@{\hspace{1cm}}r@{}l}
                   $\Prop{}:$&\,$\Type{}$ &
                   $\proof{}:$&\,$\Prop{}\arr{}\Type{}$ &
                   $\type{}:$&\,$\Type{}$ &
                   $\term{}:$&\,$\type{}\arr{}\Type{}$
                 \end{tabular}
               \end{center}
               \hspace{0.2cm}Primitive Connectives
               \begin{center}
                 \begin{tabular}{l@{\hspace{1.0cm}}l}
                   $\top:\Prop$ &
                   $\bot:\Prop$ \\
                   $\neg\arg:\Prop\arr\Prop$ &
                   $\arg\land\arg:\Prop\arr\Prop\arr\Prop$ \\
                   $\arg\lor\arg:\Prop\arr\Prop\arr\Prop$ &
                   $\arg\imp\arg:\Prop\arr\Prop\arr\Prop$ \\
                   $\arg\eqv\arg:\Prop\arr\Prop\arr\Prop$ &
                   $\forall\,\arg\,\arg:\Pi\alpha:\type.(\term~\alpha
                   \arr\Prop)\arr\Prop$ \\
                   $\forall_{\type}\,\arg:(\type\arr\Prop)\arr\Prop$ &
                   $\exists\,\arg\,\arg:\Pi\alpha:\type.(\term~\alpha
                   \arr\Prop)\arr\Prop$ \\
                   $\exists_{\type}\,\arg:(\type\arr\Prop)\arr\Prop$ &
                   $\arg=_{\arg}\arg:\Pi\alpha:\type.\term~\alpha\arr
                   \term~\alpha\arr\Prop$
                 \end{tabular}
               \end{center}
             }
           }
           \caption{\dedukti{} Declarations of \tffm{} Symbols}
           \label{fig:deeplogic1}
\end{figure}

\begin{figure}[ht!]
  \framebox[\textwidth][c]
           {\parbox{\textwidth}
             {\small
               \hspace{0.2cm}Translation Function for Types
               \begin{center}
                 \begin{tabular}{c@{\hspace{1.0cm}}c}
                   $\tr \alpha := \alpha$ &
                   $\tr {T(\tau_1,\ldots,\tau_m)} := T~{\tr {\tau_1}}~\ldots~
                   {\tr {\tau_m}}$
                 \end{tabular}
               \end{center}
               \hspace{0.2cm}Translation Function for Terms
               \begin{center}
                 \begin{tabular}{c@{\hspace{1.0cm}}c}
                   $\tr x := x$ &
                   $\tr {f(\tau_1,\ldots,\tau_m;e_1,\ldots,e_n)} :=
                   f~{\tr {\tau_1}}~\ldots~{\tr {\tau_m}}~{\tr {e_1}}~
                   \ldots~{\tr {e_n}}$
                 \end{tabular}
               \end{center}
               \hspace{0.2cm}Translation Function for Formul\ae{}
               \begin{center}
                 \begin{tabular}{r@{\hspace{1pt}}l@{\hspace{1.0cm}}
                     r@{\hspace{1pt}}l}
                   $\tr \top$ & $:= \top$ &
                   $\tr \bot$ & $:= \bot$ \\
                   $\tr {\neg \varphi}$ & $:= \neg {\tr \varphi}$ &
                   $\tr {\varphi_1 \land \varphi_2}$ & $:=
                   \tr {\varphi_1} \land \tr {\varphi_2}$ \\
                   $\tr {\varphi_1 \lor \varphi_2}$ & $:=
                   \tr {\varphi_1} \lor \tr {\varphi_2}$ &
                   $\tr {\varphi_1 \imp \varphi_2}$ & $:=
                   \tr {\varphi_1} \imp \tr {\varphi_2}$ \\
                   $\tr {\varphi_1 \eqv \varphi_2}$ & $:=
                   \tr {\varphi_1} \eqv \tr {\varphi_2}$ &
                   $\tr {e_1 =_\tau e_2}$ & $:=
                   \tr {e_1} =_{\tr \tau} \tr {e_2}$ \\
                   $\tr {\forall x : \tau.~\varphi}$ & $:=
                   \forall~\tr \tau~(\lambda x : \term~\tr \tau.~\tr \varphi)$ &
                   $\tr {\exists x : \tau.~\varphi}$ & $:=
                   \exists~\tr \tau~(\lambda x : \term~\tr \tau.~\tr \varphi)$ \\
                   $\tr {\forall_\type \alpha : \type.~\varphi}$ & $:=
                   \forall_\type~(\lambda \alpha : \type~.~\tr \varphi)$ &
                   $\tr {\exists_\type \alpha : \type.~\varphi}$ & $:=
                   \exists_\type~(\lambda \alpha : \type~.~\tr \varphi)$ \\
                   \multicolumn{4}{c}{
                     $\tr {P(\tau_1,\ldots,\tau_m;e_1,\ldots,e_n)} :=
                     P~{\tr {\tau_1}}~\ldots~{\tr {\tau_m}}~{\tr {e_1}}~
                     \ldots~{\tr {e_n}}$}
                 \end{tabular}
               \end{center}
               \hspace{0.2cm}Translation Function for Typing Contexts
               \begin{center}
                 \begin{tabular}{c@{\hspace{1.0cm}}c}
                   $\tr \emptyset := \emptyset$ &
                   $\tr {\Delta, x : \tau} := \tr \Delta, x : \term~\tr \tau$
                 \end{tabular}
               \end{center}
               \hspace{0.2cm}Translation Function for Theories
               \begin{center}
                 \begin{tabular}{r@{\hspace{1pt}}l}
                   $\tr \emptyset$ & $:= \Gamma_0 \text{ where $\Gamma_0$ is the \dedukti{} context of Fig.~\ref{fig:deeplogic1}}$ \\
                   $\tr {\T,T/m}$ & $:= \tr \T, T : \overbrace{\type \arr \ldots \arr \type}^{m \text{ times}} \arr \type $ \\
                   $\tr {\T,f:\Pi (\alpha_1, \ldots, \alpha_m).~(\tau_1, \ldots, \tau_n) \arr \tau}$ & $:= \tr \T, f : \Pi \alpha_1 : \type. \ldots \Pi \alpha_m : \type.$\\
                   & \hspace{2cm} $\term~\tr{\tau_1} \arr \ldots \arr \term~\tr{\tau_n} \arr \tr \tau$ \\
                   $\tr {\T,P:\Pi (\alpha_1, \ldots, \alpha_m).~(\tau_1, \ldots, \tau_n) \arr \omicron}$ & $:= \tr \T, P : \Pi \alpha_1 : \type. \ldots \Pi \alpha_m : \type.$\\
                   & \hspace{2cm} $\term~\tr{\tau_1} \arr \ldots \arr \term~\tr{\tau_n} \arr \Prop$ \\
                   $\tr {\T, \name : \varphi}$ & $:= \tr \T, \name : \proof~\tr \varphi$ \\
                   $\tr {\T, l \rew_\Delta r}$ & $:= \tr \T, \tr l \dkrew_\Delta \tr r$
                 \end{tabular}
               \end{center}
             }
           }
           \caption{Translation Functions from \tffm{} to \LPm{}}
           \label{fig:trfun}
\end{figure}

\subsection{From Deep to Shallow}
\label{subsec:deep_to_shallow}

Following Burel~\cite{burel2013shallow}, we add rewrite rules defining the
decoding function $\proof$ in Fig.~\ref{fig:shallowlogic1} using the
usual impredicative encoding of connectives. This transforms our deep
encoding of \tffm{} into a shallow encoding in which all connectives
are defined by the built-in constructions of \LPm{}.

This encoding is better suited for sharing proofs with other ATP
because it is less sensible to small modifications of the logic.  Any
proof found, for example, by \iproverm{} is directly usable as an
(untyped) proof in the shallow encoding.

\begin{figure}[ht!]
  \framebox[\textwidth][c]
           {\parbox{\textwidth}
             {\small
               \begin{center}
               \begin{tabular}{l@{\hspace{1pt}}l}
                 $\proof~\top$ & $\dkrew\Pi{}P:\Prop.~\proof~P\arr\proof~P$ \\
                 $\proof~\bot$ & $\dkrew\Pi{}P:\Prop.~\proof~P$ \\
                 $\proof~(\neg{}A)$ & $\dkrew\proof~A\arr\proof~\bot$ \\
                 $\proof~(A\land{}B)$ & $\dkrew\Pi{}P:\Prop.~
                 (\proof~A\arr\proof~B\arr\proof~P)\arr\proof~P$ \\
                 $\proof~(A\lor{}B)$ & $\dkrew\Pi{}P:\Prop.~
                 (\proof~A\arr\proof~P)\arr(\proof~B\arr\proof~P)
                 \arr\proof~P$ \\
                 $\proof~(A\imp{}B)$ & $\dkrew\proof~A\arr\proof~B$ \\
                 $\proof~(A\eqv{}B)$ & $\dkrew\proof~((A\imp{}B)
                 \land{}(B\imp{}A))$ \\
                 $\proof~(\forall~\tau~P)$ & $\dkrew\Pi{}x:\term~\tau.~
                 \proof~(P~x)$ \\
                 $\proof~(\forall_{\type}~P)$ & $\dkrew\Pi\alpha:\type.~
                 \proof~(P~\alpha)$ \\
                 $\proof~(\exists~\tau~P)$ & $\dkrew\Pi{}P:
                 \Prop.~(\Pi{}x:\term~\tau.~\proof~(P~x)\arr\proof~P)
                 \arr\proof~P$ \\
                 $\proof~(\exists_{\type}~P)$ & $\dkrew\Pi{}P:
                 \Prop.~(\Pi\alpha:\type.~\proof~(P~\alpha)\arr\proof~P)
                 \arr\proof~P$ \\
                 $\proof~(x=_{\tau}y)$ & $\dkrew\Pi{}P:(\term~\tau\arr
                 \Prop).~\proof~(P~x)\arr\proof~(P~y)$ \\
                 \end{tabular}
               \end{center}
             }
           }
           \caption{Shallow Definition of Logical Connectives in \dedukti{}}
           \label{fig:shallowlogic1}
\end{figure}

%% file: zenon.tex
\section{\zenm{}}
\label{sec:zenonmodulo}

\begin{figure}[hb!]
\framebox[\textwidth][c]
{\parbox{\textwidth}
{\small
\hspace{0.2cm}Closure and Quantifier-free Rules
\begin{center}
  \begin{tabular}{c@{\hspace{0.4cm}}c@{\hspace{0.4cm}}c
      @{\hspace{0.4cm}}c}
   \AxiomC {$\phantom{P}$}
   \RightLabel {$\bot$}
   \UnaryInfC  {$\Gamma,[\bot]\fCenter\bot$}
   \DisplayProof &
   \AxiomC {$\phantom{P}$}
   \RightLabel {$\neg\top$}
   \UnaryInfC  {$\Gamma,[\neg\top]\fCenter\bot$}
   \DisplayProof &
   \AxiomC {$\phantom{P}$}
   \RightLabel {$\mathrm{Ax}$}
   \UnaryInfC {$\Gamma,[P],[\neg{}P]\fCenter\bot$}
   \DisplayProof &
   \AxiomC {$\Gamma,P\fCenter\bot$}
   \AxiomC {$\Gamma,\neg{}P\fCenter\bot$}
   \RightLabel {$\mathrm{Cut}$}
   \BinaryInfC {$\Gamma\fCenter\bot$}
   \DisplayProof  \\\\
   \AxiomC {$\phantom{P}$}
   \RightLabel {$\neq$}
   \UnaryInfC {$\Gamma,[t\neq_{\tau}t]\fCenter\bot$}
   \DisplayProof &
   \AxiomC {$\phantom{P}$}
   \RightLabel  {$\mathrm{Sym}$}
   \UnaryInfC {$\Gamma,[t=_{\tau}u],[u\neq_{\tau}t]\fCenter\bot$}
   \DisplayProof &
   \Axiom $\Gamma,\neg\neg{}P,P\fCenter\bot$
   \RightLabel {$\neg\neg$}
   \UnaryInf $\Gamma,[\neg\neg{}P]\fCenter\bot$
   \DisplayProof &
   \Axiom $\Gamma,P\land{}Q,P,Q\fCenter\bot$
   \RightLabel {$\land$}
   \UnaryInf $\Gamma,[P\land{}Q]\fCenter\bot$
   \DisplayProof  \\\\
  \end{tabular}
  \begin{tabular}{c@{\hspace{0.4cm}}c}
    \Axiom $\Gamma,P\lor{}Q,P\fCenter\bot$
    \Axiom $\Gamma,P\lor{}Q,Q\fCenter\bot$
    \RightLabel {$\lor$}
    \BinaryInf $\Gamma,[P\lor{}Q]\fCenter\bot$
    \DisplayProof &
    \Axiom $\Gamma,P\imp{}Q,\neg{}P\fCenter\bot$
    \Axiom $\Gamma,P\imp{}Q,Q\fCenter\bot$
    \RightLabel {$\imp$}
    \BinaryInf $\Gamma,[P\imp{}Q]\fCenter\bot$
    \DisplayProof \\\\
    \Axiom $\Gamma,P\eqv{}Q,\neg{}P,\neg{}Q\fCenter\bot$
    \Axiom $\Gamma,P\eqv{}Q,P,Q\fCenter\bot$
    \RightLabel {$\eqv$}
    \BinaryInf $\Gamma,[P\eqv{}Q]\fCenter\bot$
    \DisplayProof &
    \Axiom $\Gamma,\neg{}(P\land{}Q),\neg{}P\fCenter\bot$
    \Axiom $\Gamma,\neg{}(P\land{}Q),\neg{}Q\fCenter\bot$
    \RightLabel {$\neg\land$}
    \BinaryInf
    $\Gamma,[\neg{}(P\land{}Q)]\fCenter\bot$
    \DisplayProof \\\\ \Axiom
    $\Gamma,\neg{}(P\lor{}Q),\neg{}P,\neg{}Q\fCenter\bot$
    \RightLabel {$\neg\lor$}
    \UnaryInf $\Gamma,[\neg{}(P\lor{}Q)]\fCenter\bot$
    \DisplayProof &
    \Axiom $\Gamma,\neg{}(P\imp{}Q),P,\neg{}Q\fCenter\bot$
    \RightLabel {$\neg\imp$}
    \UnaryInf $\Gamma,[\neg{}(P\imp{}Q)]\fCenter\bot$
    \DisplayProof \\\\
    \multicolumn{2}{c}{
      \Axiom $\Gamma,\neg{}(P\eqv{}Q),\neg{}P,Q\fCenter\bot$
      \Axiom $\Gamma,\neg{}(P\eqv{}Q),P,\neg{}Q\fCenter\bot$
      \RightLabel {$\neg\eqv$}
      \BinaryInf  $\Gamma,[\neg{}(P\eqv{}Q)]\fCenter\bot$
      \DisplayProof}
  \end{tabular}
\end{center}
} }
\caption{\llproofm{} Inference Rules of \zenm{} (part 1)}
\label{fig:rules1}
\end{figure}

\zenm{}~\cite{Zenon-Modulo} is an extension to deduction
modulo~\cite{DA03} of the first-order tableau-based automated
theorem prover \zenon{}~\cite{Zenon}. It has also been improved to
deal with typed formul\ae{} and \tff{} input files. In this paper,
we focus on the
output format of \zenm{}. After finding
a proof using its tableau-based proof-search algorithm~\cite{Zenon},
\zenon{} translates its proof tree into a \textit{low level} format called
\llproof{}, which is a classical sequent-like proof system. This format is
used for \zenon{} proofs before their automatic
translation to \coq{}.  \llproof{} is a one-sided sequent calculus
with explicit contractions in every inference rule, which is close to
an upside-down non-destructive tableau method.

\begin{figure}[ht!]
  \framebox[\textwidth][c]
    {\parbox{\textwidth} {\small
        \hspace{0.2cm}Quantifier Rules
        \begin{center}
          \begin{tabular}{c@{\hspace{0.4cm}}c@{\hspace{0.4cm}}l}
            \Axiom $\Gamma,\exists_{\type}\alpha:\type.~P(\alpha),P(\tau)
            \fCenter\bot$
            \RightLabel {$\exists_{\type}$}
            \UnaryInf $\Gamma,[\exists_{\type}\alpha:\type.~P(\alpha)]
            \fCenter\bot$
            \DisplayProof &
            \Axiom $\Gamma,\neg\forall_{\type}\alpha:\type.~P(\alpha),
            \neg{}P(\tau)\fCenter\bot$
            \RightLabel {$\neg\forall_{\type}$}
            \UnaryInf  $\Gamma,[\neg\forall_{\type}\alpha:\type.~P(\alpha)]
            \fCenter\bot$
            \DisplayProof &
            \begin{tabular}{l}
              where $\tau$ is a \\ fresh type constant
            \end{tabular} \\\\
            \Axiom $\Gamma,\forall_{\type}\alpha:\type.~P(\alpha),P(\beta)
            \fCenter\bot$
            \RightLabel {$\forall_{\type}$}
            \UnaryInf $\Gamma,[\forall_{\type}\alpha:\type.~P(\alpha)]
            \fCenter\bot$
            \DisplayProof &
            \Axiom $\Gamma,\neg\exists_{\type}\alpha:\type.~P(\alpha),
            \neg{}P(\beta)\fCenter\bot$
            \RightLabel {$\neg\exists_{\type}$}
            \UnaryInf  $\Gamma,[\neg\exists_{\type}\alpha:\type.~P(\alpha)]
            \fCenter\bot$
            \DisplayProof &
            \begin{tabular}{l}
              where $\beta$ is \\ any closed type
            \end{tabular}\\\\
            \Axiom $\Gamma,\exists{}x:\tau.~P(x),P(c)\fCenter\bot$
            \RightLabel {$\exists$}
            \UnaryInf $\Gamma,[\exists{}x:\tau.~P(x)] \fCenter\bot$
            \DisplayProof &
            \Axiom $\Gamma,\neg\forall{}x:\tau.~P(x),\neg{}P(c)\fCenter\bot$
            \RightLabel {$\neg\forall$}
            \UnaryInf $\Gamma,[\neg\forall{}x:\tau.~P(x)] \fCenter\bot$
            \DisplayProof &
            \begin{tabular}{l}
              where $c:\tau$ is a \\ fresh constant
            \end{tabular}\\\\
            \Axiom $\Gamma,\forall{}x:\tau.~P(x),P(t)\fCenter\bot$
            \RightLabel {$\forall$}
            \UnaryInf $\Gamma,[\forall{}x:\tau.~P(x)] \fCenter\bot$
            \DisplayProof &
            \Axiom $\Gamma,\neg\exists{}x:\tau.~P(x),\neg{}P(t)
            \fCenter\bot$
            \RightLabel {$\neg\exists$}
            \UnaryInf $\Gamma,[\neg\exists{}x:\tau.~P(x)] \fCenter\bot$
            \DisplayProof &
            \begin{tabular}{l}
              where $t:\tau$ is \\ any closed term
            \end{tabular}
          \end{tabular}
        \end{center}
        \hspace{0.2cm}Special Rules
        \begin{center}
          \begin{tabular}{l}
            \Axiom $\Delta,t_1\neq_{\tau^{\prime}_1}u_1\fCenter\bot$
            \AxiomC {$\ldots$}
            \Axiom
            $\Delta,t_n\neq_{\tau^{\prime}_n}u_n\fCenter\bot$
            \RightLabel {$\mathrm{Pred}$}
            \TrinaryInf
             $\Gamma,[P(\tau_1,\ldots,\tau_m;t_1,\ldots,t_n)],
                [\neg{}P(\tau_1,\ldots,\tau_m;u_1,\ldots,u_n)]
                \fCenter\bot$
                \DisplayProof \\\\
                where
                $\Delta=\Gamma\cup\{P(\tau_1,\ldots,\tau_m;t_1,
                \ldots,t_n),\neg{}P(\tau_1,\ldots,\tau_m;u_1,\ldots,u_n)\}$ \\\\
                \Axiom $\Delta,t_1\neq_{\tau^{\prime}_1}u_1\fCenter\bot$
                \AxiomC {$\ldots$}
                \Axiom  $\Delta,t_n\neq_{\tau^{\prime}_n}u_n\fCenter\bot$
                \RightLabel {$\mathrm{Fun}$}
                \TrinaryInf
                $\Gamma,[f(\tau_1,\ldots,\tau_m;t_1,\ldots,t_n)
                  \neq_{\tau}
                  f(\tau_1,\ldots,\tau_m;u_1,\ldots,u_n)]
                \fCenter\bot$ \DisplayProof \\\\ where
                $\Delta=\Gamma\cup\{f(\tau_1,\ldots,\tau_m;t_1,
                \ldots,t_n)\neq_{\tau}f(\tau_1,\ldots,\tau_m;u_1,\ldots,u_n)\}$ \\\\
                \Axiom $\Delta,H_{11},\ldots,H_{1m}\fCenter\bot$
                \AxiomC {$\ldots$}
                \Axiom  $\Delta,H_{n1},\ldots,H_{nq}\fCenter\bot$
                \RightLabel {$\mathrm{Ext}(\mathrm{name},\mathrm{args},
                  C_1,\ldots,C_p,H_{11},\ldots,H_{nq})$}
                \TrinaryInf $\Gamma,[C_1],\ldots,[C_p]\fCenter\bot$
                \DisplayProof \\\\
                where $\Delta=\Gamma\cup\{C_1,\ldots,C_p\}$
          \end{tabular}
        \end{center}
    } }
    \caption{\llproofm{} Inference Rules of \zenm{} (part 2)}
    \label{fig:rules2}
\end{figure}

We present in Figs.~\ref{fig:rules1} and~\ref{fig:rules2} the new
proof system \llproofm{}, an adaptation of \zenon{} output format
\llproof{}~\cite{Zenon} to deduction modulo and \tff{} typing.

Normalization and deduction steps may interleave anywhere in the final
proof tree. This leads to the introduction of the congruence relation
$=_{\mathcal{RE}}$ inside rules of Figs.~\ref{fig:rules1}
and~\ref{fig:rules2}: if the formula $P$ is in normal form (with
respect to $\mathcal{RE}$), we denote by $[P]$ any formula congruent
to $P$ modulo $=_{\mathcal{RE}}$.

Extension of \llproof{} to \tff{} typing leads to the introduction of
four new rules for quantification over type variables $\exists_\type$,
$\neg\forall_\type$, $\forall_\type$ and $\neg\exists_\type$, and also
to introduce some type information into other rules dealing with
equality or quantification. For instance, equality of two closed terms
$t$ and $u$, both of type $\tau$, is denoted by $t=_{\tau}u$. For
predicate and function symbols, we first list types, then terms,
separated by a semi-colon.

Finally, last difference regarding rules presented in~\cite{Zenon} is
the removal of rules ``definition'' and ``lemma''. \zenm{}, unlike
\zenon{}, does not need to explicitly unfold definitions and
the lemma constructions have been removed.

\subsection{Translation of \zenm{} Proofs into \LPm{}}
\label{subsec:translation}

\begin{figure}[ht!]
  \framebox[\textwidth][c]
           {\parbox{\textwidth}
             {\small
               \hspace{0.2cm}\zenm{} Rules
               \begin{center}
                 \begin{tabular}{l@{\hspace{1pt}}l}
                   $\mrule_{\bot}:$&$\proof~\bot\arr\seq$ \\
                   $\mrule_{\neg\top}:$&$\proof~(\neg\top)\arr\seq$ \\
                   $\mrule_{Ax}:$&$\Pi{}P:\Prop.~\proof~P\arr\proof~(\neg{}P)
                   \arr\seq$ \\
                   $\mrule_{Cut}:$&$\Pi{}P:\Prop.~(\proof~P\arr\seq)\arr(
                   \proof~(\neg{}P)\arr\seq)\arr\seq$ \\
                   $\mrule_{\neq}:$&$\Pi\alpha:\type.~\Pi{}t:\term~\alpha.~
                   \proof~(t\neq_{\alpha}t)\arr\seq$ \\
                   $\mrule_{Sym}:$&$\Pi\alpha:\type.~\Pi{}t,u:\term~\alpha.~
                   \proof~(t=_{\alpha}u)\arr\proof~(u\neq_{\alpha}t)\arr\seq$ \\
                   $\mrule_{\neg\neg}:$&$\Pi{}P:\Prop.~(\proof~P\arr\seq)\arr
                   \proof~(\neg\neg{}P)\arr\seq$ \\
                   $\mrule_{\land}:$&$\Pi{}P,Q:\Prop.~(\proof~P\arr\proof~Q\arr
                   \seq)\arr\proof~(P\land{}Q)\arr\seq$ \\
                   $\mrule_{\lor}:$&$\Pi{}P,Q:\Prop.~(\proof~P\arr\seq)\arr
                   (\proof~Q\arr\seq)\arr\proof~(P\lor{}Q)\arr\seq$ \\
                   $\mrule_{\imp}:$&$\Pi{}P,Q:\Prop.~(\proof~(\neg{}P)\arr\seq)
                   \arr(\proof~Q\arr\seq)\arr\proof~(P\imp{}Q)\arr\seq$ \\
                   $\mrule_{\eqv}:$&$\Pi{}P,Q:\Prop.~(\proof~(\neg{}P)\arr\proof~
                   (\neg{}Q)\arr\seq)\arr(\proof~P\arr\proof~Q\arr\seq)\arr
                   \proof~(P\eqv{}Q)\arr\seq$ \\
                   $\mrule_{\neg\land}:$&$\Pi{}P,Q:\Prop.~(\proof~(\neg{}P)\arr
                   \seq)\arr(\proof~(\neg{}Q)\arr\seq)\arr\proof~(\neg(P
                   \land{}Q))\arr\seq$ \\
                   $\mrule_{\neg\lor}:$&$\Pi{}P,Q:\Prop.~(\proof~(\neg{}P)
                   \arr\proof~(\neg{}Q)\arr\seq)\arr\proof~(\neg(P\lor{}Q))
                   \arr\seq$ \\
                   $\mrule_{\neg\imp}:$&$\Pi{}P,Q:\Prop.~(\proof~P\arr\proof~
                   (\neg{}Q)\arr\seq)\arr\proof~(\neg(P\imp{}Q))\arr\seq$ \\
                   $\mrule_{\neg\eqv}:$&$\Pi{}P,Q:\Prop.~(\proof~(\neg{}P)\arr
                   \proof~Q\arr\seq)\arr(\proof~P\arr\proof~(\neg{}Q)\arr\seq)
                   \arr\proof~(\neg(P\eqv{}Q))\arr\seq$ \\
                   $\mrule_{\exists}:$&$\Pi\alpha:\type.~\Pi{}P:(\term~\alpha\arr
                   \Prop).~(\Pi{}t:\term~\alpha.~(\proof~(P~t)\arr\seq))\arr
                   \proof~(\exists~\alpha~P)\arr\seq$ \\
                   $\mrule_{\forall}:$&$\Pi\alpha:\type.~\Pi{}P:(\term~\alpha
                   \arr\Prop).~\Pi{}t:\term~\alpha.~(\proof~(P~t)\arr\seq)\arr
                   \proof~(\forall~\alpha~P)\arr\seq$ \\
                   $\mrule_{\neg\exists}:$&$\Pi\alpha:\type.~\Pi{}P:(\term~
                   \alpha\arr\Prop).~\Pi{}t:\term~\alpha.~(\proof~(\neg(P~t))
                   \arr\seq)\arr
                   \proof~(\neg(\exists~\alpha~P)\arr\seq$ \\
                   $\mrule_{\neg\forall}:$&$\Pi\alpha:\type.~\Pi{}P:(\term~
                   \alpha\arr\Prop).~(\Pi{}t:\term~\alpha.~(\proof~(\neg(P~t))
                   \arr\seq))\arr
                   \proof~(\neg\forall~\alpha~P)\arr\seq$ \\
                   $\mrule_{\exists_{\type}}:$&$\Pi{}P:(\type\arr\Prop).~(\Pi
                   \alpha:\type.~(\proof~(P~\alpha)\arr\seq))\arr\proof~
                   (\exists_{\type}~P)\arr\seq$ \\
                   $\mrule_{\forall_{\type}}:$&$\Pi{}P:(\type\arr\Prop).~\Pi
                   \alpha:\type.~(\proof~(P~\alpha)\arr\seq)\arr\proof~
                   (\forall_{\type}~f)\arr\seq$ \\
                   $\mrule_{\neg\exists_{\type}}:$&$\Pi{}P:(\type\arr\Prop).~
                   \Pi\alpha:\type.~(\proof~(\neg(P~\alpha))\arr\seq)\arr
                   \proof~(\neg(\exists_{\type}~P))\arr\seq$ \\
                   $\mrule_{\neg\forall_{\type}}:$&$\Pi{}P:(\type\arr\Prop).~
                   (\Pi\alpha:\type.~(\proof~(\neg(P~\alpha))\arr\seq))\arr
                   \proof~(\neg(\forall_{\type}~P))\arr\seq$ \\
                   $\mrule_{Subst}:$&$\Pi\alpha:\type.~\Pi{}P:(\term~\alpha\arr
                   \Prop).~\Pi{}t,u:\term~\alpha.~(\proof~(t\neq_{\alpha}u)\arr
                   \seq)\arr$ \\
                   ~&\hspace*{2.3cm}$(\proof~(P~u)\arr\seq)\arr\proof~(P~t)
                   \arr\seq$
                 \end{tabular}
               \end{center}
             }
           }
           \caption{\llproofm{} in \LPm{}}
           \label{fig:deeprules1}
\end{figure}

We present in Fig.~\ref{fig:deeprules1} a deep embedding of \llproofm{}
into \LPm{}. We declare a constant for each inference rule, except
for special rules $\mathrm{Pred}$ and $\mathrm{Fun}$ which have a
dependency on the arity $n$ of their underlying predicate and function.
Fortunately, they can be expressed with the following $\mathrm{Subst}$
inference rule which corresponds to the substitution in a predicate $P$
of a subterm $t:\tau^{\prime}$ by another $u:\tau^{\prime}$:
\begin{prooftree}
  \AxiomC {$\Gamma,P(\vec{\tau};t),t\neq_{\tau^\prime}u\vdash\bot$}
  \AxiomC {$\Gamma,P(\vec{\tau};t),P(\vec{\tau};u)\vdash\bot$}
  \RightLabel {$\mathrm{Subst}$}
  \BinaryInfC {$\Gamma,P(\vec{\tau};t)\vdash\bot$}
\end{prooftree}
The special rules $\mathrm{Pred}$ and $\mathrm{Fun}$ can be easily decomposed
into $n$ applications of the $\mathrm{Subst}$ rule. For instance, for a
binary predicate $P$, from (we omit to repeat the context $\Gamma$)
\begin{prooftree}
  \AxiomC {$\Pi_1$}
  \UnaryInfC {$t_1\neq_{\tau^{\prime}}u_1\vdash\bot$}
  \AxiomC {$\Pi_2$}
  \UnaryInfC {$t_2\neq_{\tau^{\prime\prime}}u_2\vdash\bot$}
  \RightLabel {$\mathrm{Pred}$}
  \BinaryInfC {$P(\vec{\tau};t_1,t_2),\neg{}P(\vec{\tau};u_1,u_2)\vdash\bot$}
\end{prooftree}
we obtain
\begin{prooftree}
  \AxiomC {$\Pi_1$}
  \UnaryInfC {$t_1\neq_{\tau^{\prime}}u_1\vdash\bot$}
  \AxiomC {$\Pi_2$}
  \UnaryInfC {$t_2\neq_{\tau^{\prime\prime}}u_2\vdash\bot$}
  \AxiomC {$~$}
  \RightLabel {$\mathrm{Ax}$}
  \UnaryInfC {$P(\vec{\tau};u_1,u_2)$}
  \RightLabel {$\mathrm{Subst}$}
  \BinaryInfC {$P(\vec{\tau};u_1,t_2)$}
  \RightLabel {$\mathrm{Subst}$}
  \BinaryInfC {$P(\vec{\tau};t_1,t_2),\neg{}P(\vec{\tau};u_1,u_2)\vdash\bot$}
\end{prooftree}

\begin{figure}[ht!]
  \framebox[\textwidth][c]
           {\parbox{\textwidth}
             {\small
               \hspace{0.2cm}Translation Function for Sequents
               \begin{center}
                 \begin{tabular}{c}
                   $\tr {[\varphi_1],\ldots,[\varphi_n] \vdash \bot} :=
                   x_{\varphi_1} : \proof~{\tr {\varphi_1}}, \ldots, x_{\varphi_n} : \proof~
                          {\tr {\varphi_n}}$
                 \end{tabular}
               \end{center}
               \hspace{0.2cm}Translation Function for Proofs
               \begin{center}
                 \begin{tabular}{l}
                   \tr {
                     \AxiomC {$\Pi_1$}
                     \UnaryInfC {$\Delta,H_{11},\ldots,H_{1m}\vdash\bot$}
                     \AxiomC {$\ldots$}
                     \AxiomC {$\Pi_n$}
                     \UnaryInfC {$\Delta,H_{n1},\ldots,H_{nq}\vdash\bot$}
                     \RightLabel {$\mathrm{Rule(Arg_1,\ldots,Arg_r)}$}
                     \TrinaryInfC {$\Gamma,C_1,\ldots,C_p\vdash\bot$}
                     \DisplayProof
                   } \\
                   $:=$ \\
                   \begin{tabular}{r@{\hspace{4pt}}l}
                     $\mrule_{\mathrm{Rule}}$ &
                     $\tr{\mathrm{Arg}_1}\ldots\tr{\mathrm{Arg}_r}$ \\
                     &$(\lambda{}x_{H_{11}}:\proof~\tr{H_{11}}.\ldots.
                     \lambda{}x_{H_{1m}}:\proof~\tr{H}_{1m}.\tr{\Pi_1})$ \\
                     &$\vdots$ \\
                     &$(\lambda{}x_{H_{n1}}:\proof~\tr{H_{n1}}.\ldots.
                     \lambda{}x_{H_{nq}}:\proof~\tr{H_{nq}}.\tr{\Pi_n})$ \\
                     &$x_{C_1}\ldots{}x_{C_p}$
                   \end{tabular}
                 \end{tabular}
               \end{center}
             }
           }
           \caption{Translation Functions for \llproofm{} Proofs into \LPm}
           \label{fig:trproof}
\end{figure}

~\\

In Fig.~\ref{fig:trproof}, we present the translation function for
\llproofm{} sequents and proofs into \LPm{}.
Let us present a simple example. We want to translate this proof tree: \\
\begin{center}
  \begin{tabular}{rl}$\Pi :=$ & \AxiomC {$\Pi_P$} \UnaryInfC
    {$\Gamma,P\lor{}Q,P\vdash\bot$} \AxiomC {$\Pi_Q$} \UnaryInfC
    {$\Gamma,P\lor{}Q,Q\vdash\bot$} \RightLabel
    {$\lor\scriptstyle{(P,Q)}$} \BinaryInfC
    {$\Gamma, P\lor{}Q\vdash\bot$} \DisplayProof
  \end{tabular}
\end{center}
where $\Pi_P$ and $\Pi_Q$ are respectively proofs of sequents
$\Gamma,P\vdash\bot$ and $\Gamma,Q\vdash\bot$, and where we annotate rule names
with its parameters.
Then, by applying the translation procedure of Figs.~\ref{fig:trfun}
and~\ref{fig:trproof}, we obtain the \dedukti{} term
\[ \mrule_{\lor}\;\tr{P}\;\tr{Q}\;(\lambda{}x_P:\proof\;\tr{P}.
\tr{\Pi_P})\;(\lambda{}x_Q:\proof\;\tr{Q}.\tr{\Pi_Q})\;x_{P \lor Q} \]
where the notation $\tr{x}$ means the translation of $x$ into
\LPm{}, and $x_P$ is a variable declared of type $\proof\;\tr{P}$. We then
check that $\Pi$ is a proof of the sequent $\Gamma, P\lor{}Q\vdash\bot$ in a
\tffm{} theory \T, by checking that
$\tr{\T}, \tr{\Gamma, P\lor{}Q} \vdash \tr \Pi : \seq$ in \LPm{}.

More generally, for any \llproofm{} proof $\Pi$ and any sequent
$\Gamma \vdash \bot$, we check that $\Pi$ is a proof of
$\Gamma \vdash \bot$ by checking the \LPm{} typing judgment
$\tr{\T}, \tr{\Gamma} \vdash \tr \Pi : \seq$.

\subsection{Shallow Embedding of \llproofm{}}
\label{subsec:shallowllproof}
The embedding of \llproofm{} presented in Fig.~\ref{fig:deeprules1} can
also be lifted to a shallow embedding. In Fig.~\ref{fig:shallowrules}
of Appendix~\ref{sec:appnd}, we present rewrite rules that prove all
constants corresponding to \llproofm{} inference rules into the logic
presented in Sec.~\ref{sec:dedukti}. This has been written in
\dedukti{} syntax and successfully checked by \dedukti{} (see the file
\texttt{modulogic.dk} distributed with the source code of
\zenm{}\footnote{https://www.rocq.inria.fr/deducteam/ZenonModulo/}).
The only remaining axiom is the law of excluded middle. This shows
the soundness of \llproofm{} relatively to the
consistency of the logic of Sec.~\ref{sec:dedukti}.

%% file: bench.tex
\section{Experimental Results}
\label{sec:bench}

\zenm{} helps to automatically discharge proof obligations in
particular in the two projects \focalize{}~\cite{focalize} and
\bware{}~\cite{BWare}. We present in this section some examples of
theories, and simple related properties, that are handled successfully
by \zenm{}, and its translation to \dedukti{}.

\myfigure{A \tffm{} Theory of Booleans}{fig:thbool}{
  \figitem {Declarations}

  \begin{tabular}{rl}
    $\bool/0$ & \\
    $\bfalse$ & $: \bool$\\
    $\btrue$ & $: \bool$\\
    $\bnot$ & $: \bool \arr \bool$\\
    $\arg \band \arg$ & $: \bool \arr \bool \arr \bool$\\
    $\arg \bor \arg$ & $: \bool \arr \bool \arr \bool$\\
    $\ifte \arg \arg \arg \arg$ & $: \Pi \alpha.~\bool \arr \alpha \arr
    \alpha \arr \alpha$\\
  \end{tabular}

  \figitem {Rewrite rules}

  \begin{tabular}{rlrl}
    $\btrue \band a$ & $\rew a$ &                            $\btrue \bor a$ & $\rew \btrue$ \\
    $a \band \btrue$ & $\rew a$ &                            $a \bor \btrue$ & $\rew \btrue$ \\
    $\bfalse \band a$ & $\rew \bfalse$ &                     $\bfalse \bor a$ & $\rew a$ \\
    $a \band \bfalse$ & $\rew \bfalse$ &                     $a \bor \bfalse$ & $\rew a$ \\
    $a \band a$ & $\rew a$ &                                 $a \bor a$ & $\rew a$ \\
    $a \band (b \band c)$ & $\rew (a \band b) \band c$ &     $a \bor (b \bor c)$ & $\rew (a \bor b) \bor c$ \\
    & & & \\
    $\bnot \btrue$ & $\rew \bfalse$ &                        $a \band (b \bor c)$ & $\rew (a \band b) \bor (a \band c)$ \\
    $\bnot \bfalse$ & $\rew \btrue$ &                        $(a \bor b) \band c$ & $\rew (a \band c) \bor (b \band c)$ \\
    $\bnot (\bnot a)$ & $\rew a$ &                           & \\
    $\bnot (a \bor b)$ & $\rew (\bnot a) \band (\bnot b)$ &  $\ifte \alpha \btrue t e$ & $\rew t$\\
    $\bnot (a \band b)$ & $\rew (\bnot a) \bor (\bnot b)$ &  $\ifte \alpha \bfalse t e$ & $\rew e$ \\
  \end{tabular}

  \figitem {Extension deduction rules}

  \begin{prooftree}
    \AxiomC{$\Gamma,\neg P(\btrue)\vdash\bot$}
    \AxiomC{$\Gamma,\neg P(\bfalse)\vdash\bot$}
    \RightLabel{Ext(bool-case-$\neg\forall$, $P$)}
    \BIC{$\Gamma,[\neg \forall b : \bool.~P(b)]\vdash\bot$}
  \end{prooftree}
  \begin{prooftree}
    \AxiomC{$\Gamma,P(\btrue)\vdash\bot$}
    \AxiomC{$\Gamma,P(\bfalse)\vdash\bot$}
    \RightLabel{Ext(bool-case-$\exists$, $P$)}
    \BIC{$\Gamma,[\exists b : \bool.~P(b)]\vdash\bot$}
  \end{prooftree}
}

\subsection{Application to \focalize{}}
\label{subsec:focalizeapp}

\focalize{} is a framework for specifying, developing and certifying
programs. The specification language is first-order logic and proofs
can be discharged to \zenon{} or \zenm{}. The \focalize{} compiler
produces both a regular program written in \ocaml{} and a certificate
written either in \coq{} or in \dedukti{} (but only the \dedukti{}
output can be used from \zenm{}).

In \focalize{}, specifications usually rely a lot on the primitive
type $\bool$ so it is important that \zenm{} deals with booleans
efficiently. In order to prove all propositional tautologies, it is
enough to add the following rules for reasoning by case on booleans
(together with truth tables of connectives):

\begin{prooftree}
  \AxiomC{$\Gamma,\neg P(\btrue)\vdash\bot$}
  \AxiomC{$\Gamma,\neg P(\bfalse)\vdash\bot$}
  \RightLabel{Ext(bool-case-$\neg\forall$, $P$)}
  \BIC{$\Gamma,[\neg \forall b : \bool.~P(b)]\vdash\bot$}
\end{prooftree}
\begin{prooftree}
  \AxiomC{$\Gamma,P(\btrue)\vdash\bot$}
  \AxiomC{$\Gamma,P(\bfalse)\vdash\bot$}
  \RightLabel{Ext(bool-case-$\exists$, $P$)}
  \BIC{$\Gamma,[\exists b : \bool.~P(b)]\vdash\bot$}
\end{prooftree}

\begin{figure}[ht!]
\framebox[\textwidth][c]
{\parbox{\textwidth}
{\small
  \[\begin{array}{l}
      \lambda x_1 : \proof {(\neg(\forall~\bool~(\lambda x : \term~\bool.~\forall~\bool~(\lambda y : \term~\bool.~x \band y =_\bool y \band x))))}.\\
      ~~\mrule_{\text{bool-case-$\neg\forall$}}\\
      ~~~~(\lambda x : \term~\bool.~\forall~\bool~(\lambda y : \term~\bool.~x \band y =_\bool y \band x))\\
      ~~~~(\lambda x_2 : \proof (\neg(\forall~\bool~(\lambda y : \term~\bool.~y =_\bool y))).\\
      ~~~~~~\mrule_{\neg\forall}~\bool\\
      ~~~~~~~(\lambda y : \term~\bool.~y \neq_\bool y)\\
      ~~~~~~~(\lambda a : \term~\bool.\\
      ~~~~~~~~~\lambda x_3 : \proof {(a \neq_\bool a)}.\\
      ~~~~~~~~~~~\mrule_{\neq}~\bool~a~x_3)\\
      ~~~~~~~x_2)\\
      ~~~~(\lambda x_4 : \proof (\neg(\forall~\bool~(\lambda y : \term~\bool.~\bfalse =_\bool \bfalse))).\\
      ~~~~~~~\mrule_{\neg\forall}~\bool\\
      ~~~~~~~~(\lambda y : \term~\bool.~\bfalse =_\bool \bfalse)\\
      ~~~~~~~~(\lambda a : \term~\bool.\\
      ~~~~~~~~~~\lambda x_5 : \proof {(\bfalse \neq_\bool \bfalse)}.\\
      ~~~~~~~~~~~~\mrule_{\neq}~\bool~\bfalse~x_5)\\
      ~~~~~~~~x_4)\\
      ~~~~x_1
    \end{array}\]
}}
    \caption{Proof Certificate for Commutativity of Conjunction in \dedukti{}}
    \label{fig:bool-commute}
  \end{figure}

However, we get a much smaller proof-search space and smaller proofs
by adding common algebraic laws as rewrite rules. In
Fig.~\ref{fig:thbool}, we define a theory of booleans in \tffm{}.
This theory handles idempotency and associativity of conjunction and
disjunction but not commutativity because the rule
$a \band b \dkrew b \band a$ would lead to a non terminating rewrite
system; therefore, commutativity is a lemma with the following proof:
\begin{prooftree}
  \AxiomC{}
  \RightLabel{$\neq$}
  \UIC{$a \neq_\bool a \vdash\bot$}
  \RightLabel{$\neg\forall$}
  \UIC{$\neg \forall y : \bool.~y =_\bool y \vdash\bot$}
  \AxiomC{}
  \RightLabel{$\neq$}
  \UIC{$\bfalse \neq_\bool \bfalse \vdash\bot$}
  \RightLabel{$\neg\forall$}
  \UIC{$\neg \forall y : \bool.~\bfalse =_\bool \bfalse\vdash\bot$}
  \RightLabel{Ext(bool-case-$\neg\forall$)}
  \BIC{$\neg \forall x, y : \bool.~x \band y =_\bool y \band x\vdash\bot$}
\end{prooftree}
~

The translation of this proof in \dedukti{} is shown in Fig.~\ref{fig:bool-commute}.

\subsection{Application to Set Theory}
\label{subsec:appset}

The \bware{} project is an industrial research project that aims to
provide a framework to support the automated verification of proof
obligations coming from the development of industrial applications
using the \bmth{} method~\cite{B-Book}.  The \bmth{} method relies on a particular
set theory with types. In the context of the \bware{} project, this
typed set theory has been encoded into \whyml{}, the native language of
\why{}~\cite{FP13}. To call \zenm{}, \why{} translates proof
obligations and the \bmth{} theory into \tff{} format. If it succeeds
in proving the proof obligation,
\zenm{} produces a proof certificate containing both the theory and the
term, following the model presented in Fig.~\ref{fig:certif}.

\begin{figure}[ht!]
\framebox[\textwidth][c]
{\parbox{\textwidth}
{\small
  \begin{tabular}{r@{\hspace{4pt}}l}
    $\set\arg:$&$\type\arr\type$ \\
    $\arg\in_{\arg}\arg:$&$\Pi\alpha:\type.~(\term~\set\alpha)\arr
    (\term~\set\alpha)\arr\Prop$ \\
    $\arg=_{\set\arg}\arg:$&$\Pi\alpha:\type.~(\term~\set\alpha)\arr
    (\term~\set\alpha)\arr\Prop$ \\
    $\emptyset_{\arg}:$&$\Pi\alpha:\type.~(\term~\set\alpha)$ \\
    $\arg-_{\arg}\arg:$&$\Pi\alpha:\type.~(\term~\set\alpha)\arr
    (\term~\set\alpha)\arr(\term~\set\alpha)$ \\\\
    $s=_{\set\alpha}t\dkrew$ & $
    \forall~(\alpha)~(\lambda{}x:(\term~\alpha).~x\in_{\alpha}s\eqv{}
    x\in_{\alpha}t)$ \\
    $x\in_{\alpha}\emptyset_{\alpha}\dkrew$ & $
    \bot$ \\
    $x\in_{\alpha}s-_{\alpha}t\dkrew$ & $
    x\in_{\alpha}s\land{}x\not\in_{\alpha}t$ \\ \\
  $Goal:\phantom{\dkrew}$ &
  $\proof~(\neg(\forall_{\type}~(\lambda\alpha:\type.~(\forall~(\set\alpha)~
  (\lambda{}s:(\term~\set\alpha).~s-_{\alpha}s=_{\set\alpha}\emptyset_{\alpha}))))
  \arr\proof~\bot)$ \\
  $[]~Goal\dkrew\phantom{:}$ &
  $\lambda{}x_1:\proof~(\neg(\forall_{\type}~(\lambda\alpha:\type.~(\forall~
  (\set\alpha)~(\lambda{}s:(\term~\set\alpha).~s-_{\alpha}s=_{\set\alpha}
  \emptyset_{\alpha}))))).~$ \\
  & $\mrule_{\neg\forall_{\type}}~(\lambda\alpha:\type.~(\forall~(\set\alpha)~
  \lambda{}s:(\term~\set\alpha).~s-_{\alpha}s=_{\set\alpha}\emptyset_{\alpha}))~$ \\
  & $\phantom{\mrule_{\neg\forall_{\type}}~}(\lambda\tau:\type.~$ \\
  & $\phantom{\mrule_{\neg\forall_{\type}}~(}\lambda{}x_2:\proof~(\neg(\forall~
  (\set\tau)~(\lambda{}s:(\term~\set\tau).~
  s-_{\tau}s=_{\set\tau}\emptyset_{\tau}))).~$ \\
  & $\phantom{\mrule_{\neg\forall_{\type}}~(}\mrule_{\neg\forall}~(\set\tau)~$ \\
  & $\phantom{\mrule_{\neg\forall_{\type}}~(\mrule_{\neg\forall}~}(\lambda{}s:(\term~
  \set\tau).~s-_{\tau}s=_{\set\tau}\emptyset_{\tau})~$ \\
  & $\phantom{\mrule_{\neg\forall_{\type}}~(\mrule_{\neg\forall}~}(\lambda{}
  c_1:(\term~\set\tau).~$ \\
  & $\phantom{\mrule_{\neg\forall_{\type}}~(\mrule_{\neg\forall}~(}\lambda{}x_3:\proof~
  (c_1-_{\tau}c_1\neq_{\set\tau}
  \emptyset_{\tau}).~$ \\
  & $\phantom{\mrule_{\neg\forall_{\type}}~(\mrule_{\neg\forall}~(}\mrule_{\neg\forall}~
  (\tau)~$ \\
  & $\phantom{\mrule_{\neg\forall_{\type}}~(\mrule_{\neg\forall}~(\mrule_{\neg\forall}~}
  (\lambda{}x:(\term~\tau).~(x\in_{\tau}c_1-_{\tau}c_1)\eqv(x\in_{\tau}
  \emptyset_{\tau}))~$ \\
  & $\phantom{\mrule_{\neg\forall_{\type}}~(\mrule_{\neg\forall}~(\mrule_{\neg\forall}~}
  (\lambda{}c_2:(\term~\tau).~$ \\
  & $\phantom{\mrule_{\neg\forall_{\type}}~(\mrule_{\neg\forall}~(\mrule_{\neg\forall}~(}
  \lambda{}x_4:\proof~(\neg((c_2\in_{\tau}c_1-_{\tau}c_1)\eqv(c_2\in_{\tau}
  \emptyset_{\tau}))).~$ \\
  & $\phantom{\mrule_{\neg\forall_{\type}}~(\mrule_{\neg\forall}~(\mrule_{\neg\forall}~(}
  \mrule_{\neg\eqv}~(c_2\in_{\tau}c_1-_{\tau}c_1)~$ \\
  & $\phantom{\mrule_{\neg\forall_{\type}}~(\mrule_{\neg\forall}~(\mrule_{\neg\forall}~(
    \mrule_{\neg\eqv}~}(c_2\in_{\tau}\emptyset_{\tau})$ \\
  & $\phantom{\mrule_{\neg\forall_{\type}}~(\mrule_{\neg\forall}~(\mrule_{\neg\forall}~(
    \mrule_{\neg\eqv}~}(\lambda{}x_5:\proof~(\neg(c_2\in_{\tau}c_1-_{\tau}c_1)).~$ \\
  & $\phantom{\mrule_{\neg\forall_{\type}}~(\mrule_{\neg\forall}~(\mrule_{\neg\forall}~(
    \mrule_{\neg\eqv}~(}\lambda{}x_6:\proof~(c_2\in_{\tau}\emptyset_{\tau}).~$ \\
  & $\phantom{\mrule_{\neg\forall_{\type}}~(\mrule_{\neg\forall}~(\mrule_{\neg\forall}~(
    \mrule_{\neg\eqv}~(}\mrule_{\bot}~x_6)$ \\
  & $\phantom{\mrule_{\neg\forall_{\type}}~(\mrule_{\neg\forall}~(\mrule_{\neg\forall}~(
    \mrule_{\neg\eqv}~}(\lambda{}x_7:\proof~(c_2\in_{\tau}c_1-_{\tau}c_1).~$ \\
  & $\phantom{\mrule_{\neg\forall_{\type}}~(\mrule_{\neg\forall}~(\mrule_{\neg\forall}~(
    \mrule_{\neg\eqv}~(}\lambda{}x_8:\proof~(\neg(c_2\in_{\tau}\emptyset_{\tau})).~$ \\
  & $\phantom{\mrule_{\neg\forall_{\type}}~(\mrule_{\neg\forall}~(\mrule_{\neg\forall}~(
    \mrule_{\neg\eqv}~(}\mrule_{\land}~(c_2\in_{\tau}c_1)~$ \\
  & $\phantom{\mrule_{\neg\forall_{\type}}~(\mrule_{\neg\forall}~(\mrule_{\neg\forall}~(
    \mrule_{\neg\eqv}~(\mrule_{\land}~}(c_2\not\in_{\tau}c_1)~$ \\
  & $\phantom{\mrule_{\neg\forall_{\type}}~(\mrule_{\neg\forall}~(\mrule_{\neg\forall}~(
    \mrule_{\neg\eqv}~(\mrule_{\land}~}(\lambda{}x_9:\proof~(c_2\in_{\tau}c_1).~$ \\
  & $\phantom{\mrule_{\neg\forall_{\type}}~(\mrule_{\neg\forall}~(\mrule_{\neg\forall}~(
    \mrule_{\neg\eqv}~(\mrule_{\land}~(}\lambda{}x_9:\proof~(c_2\not\in_{\tau}c_1).~
  $ \\
  & $\phantom{\mrule_{\neg\forall_{\type}}~(\mrule_{\neg\forall}~(\mrule_{\neg\forall}~(
    \mrule_{\neg\eqv}~(\mrule_{\land}~(}\mrule_{Ax}~(c_2\not\in_{\tau}c_1)~$ \\
  & $\phantom{\mrule_{\neg\forall_{\type}}~(\mrule_{\neg\forall}~(\mrule_{\neg\forall}~(
    \mrule_{\neg\eqv}~(\mrule_{\land}~(\mrule_{Ax}~}x_8$ \\
  & $\phantom{\mrule_{\neg\forall_{\type}}~(\mrule_{\neg\forall}~(\mrule_{\neg\forall}~(
    \mrule_{\neg\eqv}~(\mrule_{\land}~(\mrule_{Ax}~}x_9)$ \\
  & $\phantom{\mrule_{\neg\forall_{\type}}~(\mrule_{\neg\forall}~(\mrule_{\neg\forall}~(
    \mrule_{\neg\eqv}~(\mrule_{\land}~(}x_7)$ \\
  & $\phantom{\mrule_{\neg\forall_{\type}}~(\mrule_{\neg\forall}~(\mrule_{\neg\forall}~(
    \mrule_{\neg\eqv}~(}x_4)$ \\
  & $\phantom{\mrule_{\neg\forall_{\type}}~(\mrule_{\neg\forall}~(\mrule_{\neg\forall}~(}
  x_3)$ \\
  & $\phantom{\mrule_{\neg\forall_{\type}}~(\mrule_{\neg\forall}~(}x_2)$ \\
  & $\phantom{\mrule_{\neg\forall_{\type}}~(}x_1)$
\end{tabular}
}}
\caption{Proof Certificate for a \bmth{} Set Theory Property in \dedukti{}}
\label{fig:certif}
\end{figure}

The \bware{} project provides a large benchmark made of 12,876 proof
obligations coming from industrial projects. The embedding presented in
this paper allowed us to verify with \dedukti{} all the 10,340 proof
obligations that are proved by \zenm{}.

Let us present a small subset of this set theory, and a simple example
of \llproofm{} proof produced by \zenm{}. The theory consists of three
axioms that have been turned into rewrite rules. We define constructors:
a type constructor $\set$, the membership
predicate $\in$, equality on sets $=_{\set}$, the empty set
$\emptyset$ and difference of sets $-$. For readability, we
use an infix notation and let type parameters of functions and
predicates in subscript. We want to prove the property
\[ \forall_{\type}\alpha:\type.~\forall{}s:\set\alpha.~s-_{\alpha}s
=_{\set\alpha}\emptyset_{\alpha} \] \clearpage
\noindent{}with the theory: \\

\begin{tabular}{c@{\hspace{2.0cm}}c}
  \begin{tabular}{r@{\hspace{1pt}}l}
    $\set\arg~$&$/1$ \\
    $\arg\in_{\arg}\arg:$&$\Pi\alpha.~\set\alpha\arr\set\alpha\arr\Prop$ \\
    $\arg=_{\set\arg}\arg:$&$\Pi\alpha.~\set\alpha\arr\set\alpha\arr\Prop$ \\
    $\emptyset_{\arg}:$&$\Pi\alpha.~\set\alpha$ \\
    $\arg-_{\arg}\arg:$&$\Pi\alpha.~\set\alpha\arr\set\alpha\arr\set\alpha$
  \end{tabular} &
  \begin{tabular}{r@{\hspace{1pt}}l}
    $s=_{\set\alpha}t\rew$ &
    $\forall{}x:\alpha.~x\in_{\alpha}s\eqv{}x\in_{\alpha}t$ \\
    $x\in_{\alpha}\emptyset_{\alpha}\rew$ &
    $\bot$ \\
    $x\in_{\alpha}s-_{\alpha}t\rew$ &
    $x\in_{\alpha}s\land{}x\not\in_{\alpha}t$
  \end{tabular} \\ &
\end{tabular} \\
The \llproof{} proof tree generated by \zenm{} is (we omit to repeat context
$\Gamma$):
\begin{prooftree}
  \AxiomC {$\phantom{P}$}
  \RightLabel {$\bot$}
  \UnaryInf $\neg(c_2\in_{\tau}c_1-_{\tau}c_1),c_2\in_{\tau}\emptyset_{\tau}
  \fCenter\bot$
  \AxiomC {$\phantom{P}$}
  \RightLabel {$\mathrm{Ax}$}
  \UnaryInf $c_2\in_{\tau}c_1,c_2\not\in_{\tau}c_1\fCenter\bot$
  \RightLabel {$\land$}
  \UnaryInf $c_2\in_{\tau}c_1-_{\tau}c_1,\neg(c_2\in_{\tau}\emptyset_{\tau})
  \fCenter\bot$
  \RightLabel {$\neg\eqv$}
  \BinaryInf $\neg((c_2\in_{\tau}c_1-_{\tau}c_1)\eqv{}(c_2\in_{\tau}\emptyset_{\tau}))
  \fCenter\bot$
  \RightLabel {$\neg\forall$}
  \UnaryInf $\neg(c_1-_{\tau}c_1=_{\set\tau}\emptyset_{\tau})\fCenter\bot$
  \RightLabel {$\neg\forall$}
  \UnaryInf $\neg(\forall{}s:\set\tau.~s-_{\tau}s=_{\set\tau}\emptyset_{\tau})
  \fCenter\bot$
  \RightLabel {$\neg\forall_{\type}$}
  \UnaryInf $\neg(\forall_{\type}\alpha:\type.~\forall{}s:\set\alpha.~
  s-_{\alpha}s=_{\set\alpha}\emptyset_{\alpha})\fCenter\bot$
\end{prooftree}
~ \\

We obtain the proof certificate of Fig.~\ref{fig:certif}
checkable by \dedukti{}, using the file
\texttt{modulogic.dk}, and that is successfully checked.

%% file: concl.tex
\section{Conclusion}

We have presented a shallow embedding of \zenm{} proofs into
\dedukti{}. For this encoding, we have needed to embed into \LPm{} an
extension to deduction modulo of the underlying logic of the \tff{}
format, denoted by \tffm{}.  We then defined \llproofm{}, the
extension to \tffm{} of the proof system \llproof{}, which is the
output format of \zenm{}. Finally, we have embedded \llproofm{} into
\LPm{} by giving the translation function for proofs. This embedding is
shallow in the sense that we have
reused the features of the target language and have not declared new
constants for connectives and inference rules. The only axiom that we
have added is the law of excluded middle.

This embedding has helped us to verify a large set of proof
obligations coming from two different projects. \focalize{} can now
benefit from deduction modulo to improve program verification when
dealing with theories. In \bware{}, this work allowed us to certify
all the proofs generated by \zenm{}.

Our work is closely related to the embedding of \iproverm{} proofs
into \dedukti{}~\cite{burel2013shallow}. The two main differences are
the assumption of the excluded middle and the extension of the logic
to deal with ML-style polymorphism. Because these shallow encodings
are close, we could easily share proofs of untyped formul\ae{} with
\iproverm{}.

We do not have to trust the full implementation of \zenm{} but only
the translation of \tffm{} problems to \LPm{} discussed in
Sec.~\ref{sec:dedukti}
and, of course, \dedukti{}. In the case of \focalize{}, we go even further
by using an external translator, \focalide{}~\cite{focalide}. Hence
\zenm{} requires no confidence in that context. As future work, we want export
this model. To achieve that, deduction tools must be able to read
\dedukti{} in addition to write some. This model improves the
confidence on automated deduction tools because it is no more possible
to introduce inconsistency inside a proof certificate. In addition, in
case of the verification of several formul\ae{}, it should be possible to
inject terms coming from different tools inside the same \dedukti{}
file. A first experiment with \zenm{} and \iproverm{} in \focalize{}
would be an interesting proof of concept.

%% file: appnd.tex
\section{Appendix: Shallow Embedding of \llproofm{} System into \dedukti{}}
\label{sec:appnd}

\begin{figure}[ht!]
  \framebox[\textwidth][c]
           {\parbox{\textwidth}
             {\footnotesize
               \hspace{0.2cm}Law of Excluded Middle and Lemmas
               \begin{center}
                 \begin{tabular}{l}
                   $ExMid(P:\Prop):\Pi{}Z:\Prop.~(\proof~P\arr\proof~Z)\arr
                   (\proof~(\neg{}P)\arr\proof~Z)\arr\proof~Z$\\
                   $NNPP(P:\Prop):\proof~(\neg\neg{}P)\arr\proof~P$ \\
                   \hspace{0.2cm} $:=\lambda{}H_1:\proof~(\neg\neg{}P).~
                   ExMid~P~P~(\lambda{}H_2:\proof~P.~H_2)~(\lambda{}H_3:\proof~(\neg{}P).~
                   H_1~H_3~P)~~\square$ \\
                   $Contr(P:\Prop,Q:\Prop):(\proof~(P\imp{}Q)\arr\proof~(\neg{}Q
                   \arr\neg{}P))$ \\
                   \hspace{0.2cm} $:=\lambda~H_1:\proof~(P\imp{}Q).~\lambda{}H_2:\proof~(
                   \neg{}Q).~\lambda{}H_3:\proof~P.~H_2~(H_1~H_3)~~\square$ \\
                 \end{tabular}
               \end{center}
               \hspace{0.2cm}\llproof{} Inference Rules
               \begin{center}
                 \begin{tabular}{l}
                   $[]~\mrule_{\bot}\dkrew{}\lambda{}H:\proof~\bot.~H~~\square$ \\
                   $[]~\mrule_{\neg\top}\dkrew{}\lambda{}H_1:\proof~(\neg\top).~
                   H_1~(\lambda{}Z:\Prop.~\lambda{}H_2:\proof~Z.~H_2)~~\square$ \\
                   $[P:\Prop]~\mrule_{Ax}~P\dkrew{}\lambda{}H_1:\proof~P.~
                   \lambda{}H_2:\proof~(\neg{}P).~H_2~H_1~~\square$ \\
                   $[\alpha:\type,t:\term~\alpha]~\mrule_{\neq}~\alpha~t
                   \dkrew{}\lambda{}H_1:\proof~(t\neq_{\alpha}t).~
                   H_1~(\lambda{}z:(\term~\alpha\arr\Prop).~\lambda{}H_2:\proof~
                   (z~t).~H_2)~~\square$ \\
                   $[\alpha:\type,t:\term~\alpha,u:\term~\alpha]~\mrule_{Sym}~
                   \alpha~t~u\dkrew{}\lambda{}H_1:\proof~(t=_{\alpha}u).~
                   \lambda{}H_2:\proof~(u\neq_{\alpha}t).~H_2~$ \\
                   \hspace{0.6cm} $(\lambda{}z:(\term~
                   \alpha\arr\Prop).~\lambda{}H_3:\proof~(z~u).~H_1~
                   (\lambda{}x:\term~
                   \alpha.~(z~x)\imp(z~t))~(\lambda{}H_4:\proof~(z~t).~H_4)~
                   H_3)~~\square$ \\
                   $[P:\Prop]~\mrule_{Cut}~P\dkrew{}\lambda{}H_1:(\proof~P\arr
                   \proof~\bot).~\lambda{}H_2:(\proof~(\neg{}P)\arr\proof~
                   \bot).~H_2~H_1~~\square$ \\
                   $[P:\Prop]~\mrule_{\neg\neg}~P\dkrew{}\lambda{}H_1:(\proof~
                   P\arr\proof~\bot).~\lambda{}H_2:\proof~(\neg\neg{}P).~
                   H_2~H_1~~\square$ \\
                   $[P:\Prop,Q:\Prop]~\mrule_{\land}~P~Q\dkrew{}\lambda{}
                   H_1:(\proof~P\arr\proof~Q\arr\proof\bot).~\lambda{}H_2:
                   \proof~(P\land{}Q).~H_2~\bot~H_1~~\square$ \\
                   $[P:\Prop,Q:\Prop]~\mrule_{\lor}~P~Q\dkrew{}\lambda{}H_1:
                   (\proof~P\arr\proof\bot).~
                   \lambda{}H_2:(\proof~Q\arr\proof~\bot).~\lambda{}H_3:\proof~
                   (P\lor{}Q).~H_3~\bot~H_1~H_2~~\square$ \\
                   $[P:\Prop,Q:\Prop]~\mrule_{\imp}~P~Q\dkrew{}\lambda{}
                   H_1:(\proof~(\neg{}P)\arr
                   \proof~\bot).~\lambda{}H_2:(\proof~Q\arr\proof~\bot).~\lambda{}
                   H_3:\proof~(P\imp{}Q).~H_1~$ \\
                   \hspace{0.6cm} $(Contr~P~Q~H_3~H_2)~~\square$ \\
                   $[P:\Prop,Q:\Prop]~\mrule_{\eqv}~P~Q\dkrew{}\lambda{}
                   H_1:(\proof~(\neg{}P)\arr
                   \proof~(\neg{}Q)\arr\proof~\bot).~\lambda{}H_2:(\proof~P\arr
                   \proof~Q\arr\proof~\bot).~$ \\
                   \hspace{0.6cm} $\lambda{}H_3:\proof~(P\eqv{}Q).~
                   H_3~\bot~(\lambda{}H_4:(
                   \proof~P\arr\proof~Q).~\lambda{}H_5:(\proof~Q\arr\proof~P).~
                   (H_1~(Contr~P~Q~H_4~$ \\
                   \hspace{0.6cm} $(\lambda{}H_6:\proof~Q.~
                   (H_2~(H_5~H_6))~H_6)))~(\lambda{}H_7:\proof~Q.~(H_2~(H_5~H_7))~
                   H_7))~~\square$ \\
                   $[P:\Prop,Q:\Prop]~\mrule_{\neg\land}~P~Q
                   \dkrew{}\lambda{}H_1:(\proof~(\neg{}P)\arr
                   \proof~\bot).~\lambda{}H_2:(\proof~(\neg{}Q)\arr\proof~\bot).~
                   \lambda{}H_3:\proof~(\neg(P\land{}Q)).~$ \\
                   \hspace{0.6cm} $H_1~(\lambda{}H_5:
                   \proof~P.~H_2~(\lambda{}H_6:\proof~
                   Q.~H_3~(\lambda{}Z:\Prop.~\lambda{}H_4:(\proof~P\arr\proof~Q
                   \arr\proof~Z).~H_4~H_5~H_6)))~~\square$ \\
                   $[P:\Prop,Q:\Prop]~\mrule_{\neg\lor}~P~Q
                   \dkrew{}\lambda{}H_1:(\proof~(\neg{}P)\proof~
                   (\neg{}Q)\arr\proof~\bot).~\lambda{}H_2:\proof~(\neg(P\lor{}
                   Q)).~H_1~(Contr~P~(P\lor{}Q)~$ \\
                   \hspace{0.6cm} $(\lambda{}H_3:\proof~P.~
                   \lambda{}Z:\Prop.~\lambda{}H_4:(\proof~P\arr\proof~Z).~
                   \lambda{}
                   H_5:(\proof~P\arr\proof~Z).~H_4~H_3)~H_2)~(Contr~Q~(P\lor{}
                   Q)~$ \\
                   \hspace{0.6cm} $(\lambda{}H_6:\proof~Q.~\lambda{}Z:\Prop.~
                   \lambda{}H_7:(\proof~P\arr\proof~Z).~\lambda{}H_8:(\proof~Q\arr
                   \proof~Z).~H_8~H_6)~H_2)~~\square$ \\
                   $[P:\Prop,Q:\Prop]~\mrule_{\neg\imp}~P~Q\dkrew{}\lambda{}
                   H_1:(\proof~P\arr\proof~
                   (\neg{}Q)\arr\proof~\bot).~\lambda{}H_2:\proof~(\neg(P\imp{}
                   Q)).~H_2~(\lambda{}H_3:\proof~P.~$ \\
                   \hspace{0.6cm} $(H_1~H_3)~(\lambda{}H_4:
                   \proof~Q.~
                   H_2~(\lambda{}H_5:\proof~P.~H_4))~Q)~~\square$ \\
                   $[P:\Prop,Q:\Prop]~\mrule_{\neg\eqv}~P~Q\dkrew{}\lambda{}
                   H_1:(\proof~(\neg{}P)\arr
                   \proof~(\neg{}Q)).~\lambda{}H_2:(\proof~P\arr\proof~(\neg{}
                   \neg{}Q)).~$ \\
                   \hspace{0.6cm} $\lambda{}H_3:\proof~(\neg(P\eqv{}Q)).~
                   (\lambda{}H_4:\proof~
                   (\neg{}P).~H_3~(\lambda{}Z:\Prop.~\lambda{}H_5:(\proof~(P
                   \imp{}Q)\arr\proof~(Q\imp{}P)\arr\proof~Z).~$ \\
                   \hspace{0.6cm} $H_5~
                   (\lambda{}H_6:\proof~P.~
                   H_4~H_6~Q)~(\lambda{}H_7:\proof~Q.~H_1~H_4~H_7~P)))~(\lambda{}
                   H_8:\proof~P.~H_2~H_8~(\lambda{}H_9:\proof~Q.~H_3~
                   (\lambda{}Z:\Prop.~$ \\
                   \hspace{0.6cm} $\lambda{}H_{10}:(\proof~(P\imp{}Q)\arr
                   \proof~(Q\imp{}P)\arr
                   \proof~Z).~H_{10}~(\lambda{}H_{11}:\proof~P.~H_9)~
                   (\lambda{}H_{12}:\proof~Q.~
                   H_8))))~~\square$ \\
                   $[\alpha:\type,P:\term~\alpha\arr\Prop]~\mrule_{\exists}~
                   \alpha~P\dkrew{}\lambda{}H_1:(t:\term~\alpha\arr
                   \proof~(P~t)\arr\proof~\bot).~\lambda{}H_2:\proof~(\exists~
                   \alpha~P).~H_2~\bot~H_1~\square$ \\
                   $[\alpha:\type,P:\term~\alpha\arr\Prop,t:\term~\alpha]~
                   \mrule_{\forall}~\alpha~P~t
                   \dkrew{}\lambda{}H_1:(\proof~(P~t)\arr\proof~
                   \bot).~\lambda{}H_2:\proof~(\forall~\alpha~P).~H_1~(H_2~
                   t)~~\square$ \\
                   $[\alpha:\type,P:\term~\alpha\arr\Prop,t:\term~\alpha]~
                   \mrule_{\neg\exists}~\alpha~P~t
                   \dkrew{}\lambda{}H_1:(\proof~(\neg(P~t))
                   \arr\proof~\bot).~\lambda{}H_2:\proof~(\neg(\exists~\alpha~
                   P)).~H_1~$ \\
                   \hspace{0.6cm} $(\lambda{}H_4:\proof~(P~t).~H_2~(\lambda{}Z:
                   \Prop.~\lambda{}H_3:(x:\term~
                   \alpha\arr\proof~(P~x)\arr\proof~Z).~H_3~t~H_4))~~\square$ \\
                   $[\alpha:\type,P:\term~\alpha\arr\Prop]~\mrule_{\neg\forall}~
                   \alpha~P\dkrew{}\lambda{}H_1:(t:\term~\alpha\arr
                   \proof~(\neg(P~t))\arr\proof~\bot).~\lambda{}H_2:\proof~(\neg
                   (\forall~\alpha~P)).~$ \\
                   \hspace{0.6cm} $H_2~(\lambda{}t:\term~\alpha.~
                   NNPP~(P~t)~(H_1~t))~~\square$ \\
                   $[P:\type\arr\Prop]~\mrule_{\exists_{\type}}~P
                   \dkrew{}\lambda{}H_1:(\alpha:\type\arr
                   \proof~(P~\alpha)\arr\proof~\bot).~\lambda{}H_2:\proof~
                   (\exists_{\type}~P).~H_2~\bot~H_1~~\square$ \\
                   $[P:\type\arr\Prop,\alpha:\type]~\mrule_{\forall_{\type}}~P~
                   \alpha\dkrew{}\lambda{}H_1:(\proof~(P~\alpha)\arr
                   \proof~\bot).~\lambda{}H_2:\proof~(\forall_{\type}~P).~H_1~
                   (H_2~\alpha)~~\square$ \\
                   $[P:\type\arr\Prop,\alpha:\type]~\mrule_{\neg\exists_{\type}}~P~
                   \alpha\dkrew{}\lambda{}H_1:(\proof~(\neg(P~\alpha))
                   \arr\proof~\bot).~\lambda{}H_2:\proof~(\neg(\exists_{\type}~
                   P)).~H_1~$ \\
                   \hspace{0.6cm} $(\lambda{}H_4:\proof~(P~\alpha).~H_2~
                   \lambda{}Z:
                   \Prop.~\lambda{}H_3:(\beta:\type\arr
                   \proof~(P~\beta)\arr\proof~Z).~H_3~\alpha~H_4)~~\square$ \\
                   $[P:\type\arr\Prop]~\mrule_{\neg\forall_{\type}}~P
                   \dkrew{}\lambda{}H_1:(\alpha:\type\arr\proof~
                   (\neg(P~\alpha))\arr\proof~\bot).~\lambda{}H_2:\proof~(\neg
                   (\forall_{\type}~P)).~H_2~$ \\
                   \hspace{0.6cm} $(\lambda{}\alpha:\type.~
                   NNPP~(P~\alpha)~(H_1~
                   \alpha))~~\square$ \\
                   $[\alpha:\type,P:\term~\alpha\arr\Prop,t_1:\term~\alpha,
                     t_2:\term~\alpha]~\mrule_{Subst}~\alpha~P~t_1~t_2
                   \dkrew{}\lambda{}H_1:(\proof~(t_1\neq_{\alpha}
                   t_2)\arr\proof~\bot).~$ \\
                   \hspace{0.6cm} $\lambda{}H_2:(\proof~(P~t_2)\arr\proof~
                   \bot).~\lambda{}H_3:\proof~(P~t_1).~H_1~
                   (\lambda{}H_4:\proof~
                   (t_1\neq_{\alpha}t_2).~H_2~(H_4~P~H_3))~~\square$ \\
                 \end{tabular}
               \end{center}
             }
           }
           \caption{Shallow Embedding of \llproof{} into \dedukti{}}
           \label{fig:shallowrules}
\end{figure}

The deep embedding of \llproofm{} presented in
Sec.~\ref{subsec:translation} is well-typed with respect to the deep
embedding of typed deduction modulo presented in
Sec.~\ref{subsec:deepembedding}. Using the shallow embedding
presented in Sec.~\ref{subsec:deep_to_shallow}, we
can prove all the rules declared in Fig.~\ref{fig:deeprules1} by
rewriting the $\mrule_{\textsf{rule}}$ symbols using only one axiom:
the law of excluded middle. These proofs are listed in
Fig.~\ref{fig:shallowrules} where the $\square$ symbol is used to
delimit proofs.